\documentclass[12pt]{article}

\newcommand {\beq} {\begin{equation}}
\newcommand {\eeq} {\end{equation}}
 \newcommand {\ber}{\begin{eqnarray*}}
 \newcommand {\eer} {\end{eqnarray*}}
\newcommand {\bea}{\begin{eqnarray}}
 \newcommand {\eea} {\end{eqnarray}}
\newcommand{\be}{\begin{equation}}
\newcommand{\ee}{\end{equation}}
\newcommand{\eq}[1]{(\ref{#1})}

\def\ah{\hat{a}}
\def\bh{\hat{b}}

\def\ph{\hat{p}}
\def\thh{\hat{\th}}
\def\lh{\hat{\l}}

\newcommand{\ket}[1]{|#1 \rangle}
\def\rr{{\rm r}} \def\rs{{\rm s}}\def\rv{{\rm v}}

\def\a{\alpha}          
\def\b{\beta}           
\def\c{\gamma}    
\def\d{\delta}    
\def\e{\epsilon}

\def\l{\lambda} \def\L{\Lambda}
\def\m{\mu}

\def\th{\theta}

\begin{document}
 
\hfill{LPTENS-02/59}

\hfill{LPTHE-02-56}

\vspace{20pt}

\begin{center}

{\LARGE \bf String interactions and discrete \\ symmetries of the
pp--wave background}
\vspace{30pt}

{\bf Chong-Sun Chu$^{a,}$\footnote{on leave of absence from
University of Durham, UK},
Michela Petrini$^{b}$, Rodolfo Russo$^{c}$,
Alessandro Tanzini$^{d}$
}

\vspace{15pt}
{\small \em
\begin{itemize}
\item[$^a$]
Department of Physics, National Tsing Hua University,
Hsinchu, Taiwan 300, R.O.C.
\item[$^b$]
Centre de Physique Th{\'e}orique, Ecole
Polytechnique\footnote{Unit{\'e} mixte du CNRS et de l'EP, UMR
7644}, 91128 Palaiseau Cedex, France
\item[$^c$]
Laboratoire de Physique Th{\'e}orique de
l'Ecole Normale Sup{\'e}rieure,  \\
24 rue Lhomond, {}F-75231 Paris Cedex 05, France
\item[$^d$]
LPTHE, Universit{\'e} de Paris~VI-VII, 4 Place Jussieu 75252 Paris Cedex 05,
France
\end{itemize}
}

\vskip .1in {\small \sffamily chong-sun.chu@durham.ac.uk,
petrini@cpht.polytechnique.fr,
\\rodolfo.russo@lpt.ens.fr,tanzini@lpthe.jussieu.fr}

\end{center}
\begin{abstract}
Free string theory on the plane--wave background displays a discrete
$Z_2$ symmetry exchanging the two transverse $SO(4)$ rotation
groups. This symmetry should be respected also at the interacting
level. We show that the zero mode structure proposed in hep-th/0208148
can be completed to a full kinematical vertex, contrary to claims
appeared in the previous literature. We also comment on the relation
with recent works on the string--bit formalism and on the comparison
with the field theory side of the correspondence.
\end{abstract}
 
\section{Introduction} 

One of the main obstacles to a complete understanding of AdS/CFT
duality~\cite{Maldacena:1997re} has been the lack of control on the
string side of the correspondence. The presence of non--trivial R--R
form and of a curved metric in the background makes the analysis of the
string theory challenging already at the classical level. In fact most
of the achievements made in the AdS/CFT duality are restricted to the
supergravity limit ($R^4/\alpha'^2 \to\infty$), where the bulk theory
becomes more tractable. However, in this regime, the field theory side
is in a strong coupling limit ($\lambda = g_{YM}^2 N \to\infty$).
Thus, one can really check the connection between the two sides of the
duality only for those quantities that are protected by some symmetry
against quantum corrections. Because of this restriction one cannot
really probe the dynamical and most interesting aspects of the
duality.

However, in this last year significant progress has been made to lift
this restriction, at least for a very particular sector of the
original AdS/CFT correspondence. Berenstein, Maldacena and Nastase
in~\cite{Berenstein:2002jq} proposed a duality relation between a
sector of large $R$--charge operators of ${\cal N}=4$ SYM and type
IIB string theory on the maximally supersymmetric
background (plane wave)~\cite{Blau:2001ne}
\bea \label{pwave-back}
&& g_{+-} = -2,\quad g_{++} = - \mu^2 \sum_{I=1}^8 x_I x^I, \quad g_{IJ} =
\delta_{IJ}~,~ I,J=1,\ldots,8~~,   \nonumber \\ 
&&F_{+1234} = F_{+5678} = 2\mu~,~~ \phi = {\rm constant}~. 
\eea
This proposal is very interesting for two reasons. First, it has been
shown~\cite{Metsaev:2001bj} that IIB string theory in the 
plane--wave background is solvable in the light--cone gauge using the
Green-Schwarz formalism. This gives a concrete tool to study a string
theory on a non--flat background that captures (part of) the dynamics
of a gauge theory. A second important point is that the
background~\eq{pwave-back} is actually the
Penrose--limit~\cite{penrose} of the usual $AdS_5 \times S_5$
geometry. This means that the duality proposed
in~\cite{Berenstein:2002jq} is just the restriction of the original
Maldacena duality~\cite{Maldacena:1997re} to a particular sector. Thus
what we have learned so far on the AdS/CFT duality should apply also
in this new contest.

In this spirit, the proposal of~\cite{Berenstein:2002jq} provides an
interesting setup, where we can try to answer to some long-standing
questions in the AdS/CFT duality. The two main problems are: 1) how to
build an exact dictionary between states on the string side and
gauge invariant operators on the Yang--Mills side, 2) how to isolate the
relevant field theory dynamical quantities and to connect them to
string amplitudes. From the very recent
literature~\cite{Beisert:2002bb,Gross:2002mh,Vaman:2002ka,Janik:2002bd,Gomis:2002wi}
it seems that the two problems are strictly related. If we restrict to
the BMN sector of the AdS/CFT duality, we have in principle the
necessary technical tools to answer to the above questions and this
may lead to interesting progress also in the understanding of the full
AdS/CFT duality. Here, we will focus on the study of $3$--string vertex for
the background of eq.~\eq{pwave-back} and, in particular, we will discuss
how to implement all the kinematical symmetries of the background at the
level of string interactions.

In Section~2, we begin by discussing the bosonic symmetries of the
solution~\eq{pwave-back}. In particular, we focus on a discrete $Z_2$
symmetry and study its realization at the level of the string spectrum
and interaction. As remarked in~\cite{Chu:2002eu}, the presence of
this discrete symmetry in~\eq{pwave-back} gives an important physical
input in determining the string amplitudes. We then briefly recall
the techniques invented in~\cite{Cremmer:1974jq} and 
subsequently generalized
to the superstring case in~\cite{Green:1982tc,Green:hw} (see also
Chap. 11 of \cite{Green:mn} and references therein), since our
approach is based on those results. We show how to modify the
$3$--vertex construction in order to accommodate the discrete parity
in the interaction, thus providing an explicit counterexample to the
no--go theorem presented in the Appendix~C
of~\cite{Pankiewicz:2002gs}.

In the discussion of Section~3, we briefly comment on the general
symmetry properties of the string amplitudes and on their relevance
for the comparison with the results of field
theory~\cite{Beisert:2002bb,Gross:2002mh,Janik:2002bd,Gomis:2002wi} or
those derived within the string bits approach~\cite{Vaman:2002ka}. We
believe that these symmetry properties are of crucial importance in
fixing the exact dictionary between the string and gauge theory
sides. In fact, symmetry arguments usually provide, to the duality
under consideration, robust information that are less dependent on the
technical details of the computations. Therefore symmetries should be
the first thing to be checked among the various dual descriptions, 
as it was done in the original AdS/CFT proposal~\cite{Maldacena:1997re}.

\section{String theory in pp--wave background}

\subsection{Symmetries of the pp--wave string theory}

The bosonic symmetries of the IIB solution~\eq{pwave-back} that are
manifest in the light--cone gauge are summarized by the group $SO(4)
\times SO(4) \times Z_2$. The two $SO(4)$ rotate $x^i$, $i=1,2,3,4$,
and $x^{i'}$, $i'=5,6,7,8$ respectively, while the discrete $Z_2$
symmetry swaps the first and the second $SO(4)$ factors
\be\label{Z2}
Z_2 : \;\; (x_1,x_2,x_3,x_4) ~\leftrightarrow~ (x_5,x_6,x_7,x_8)~.
\ee
In addition to the bosonic symmetries, the background also preserves
32 supersymmetries. As usual, we require that string theory respects
all the bosonic and fermionic symmetries of the background. Here we
will focus in particular on the bosonic $Z_2$ symmetry, since this
point has not been well appreciated in the literature so far.

The $Z_2$ transformation~\eq{Z2} is just a particular element of the
$SO(8)$ rotation group. It is straightforward to derive its effect on the 
eight Majorana--Weyl
fermionic coordinates that survive the light--cone constraint.
In a convenient representation of the $SO(8)$
$\c$-matrices, the $Z_2$ action on the spinors 
is~\cite{Chu:2002eu}:
\be \label{theta-flip}
\theta^3 \leftrightarrow \theta^4, \quad \mbox{and} \quad \theta^7
\leftrightarrow -\theta^8,
\ee
while all the other components are unchanged. It is instructive to review how
the $Z_2$ symmetry of the background is realized in the string Hilbert
space. Having displayed how $Z_2$ acts on the fields through~\eq{Z2}
and~\eq{theta-flip}, one needs to specify how $Z_2$ acts on the states
of the Hilbert spaces. This can be achieved by specifying the action
on the ground state. Two states $\ket{0}$ and $\ket{\rv}$ play a
particular role.  The state $|\rv \rangle$, defined by\footnote{Here
and in the following we use the same conventions
as~\cite{Pankiewicz:2002gs} both for the string mode expansions and
for the matrices entering the kinematical constraints. 
Also we take $\a'=2$ below.}
\be \label{def-v}
a_n |\rv \rangle = b_n |\rv \rangle =0 \quad \forall n,
\ee
has the minimal light--cone energy (zero) and is the true vacuum state
of the theory. The state $\ket{0}$ is defined by
\be \label{def-0}
a_n |0\rangle =0, \forall n,
\quad b_n |0\rangle =0, n \neq 0, \quad \th_0 |0\rangle =0,
\ee 
and is not the state of minimal light--cone energy, since 
it has  energy $4 \mu$. In the limit $\mu \to 0$ both these states
have vanishing energy and the same is true for all the states created
by means of fermionic zero modes. In flat--space $\ket{0}$ is taken
to be true vacuum, since its definition preserves the $SO(8)$
invariance of the theory. In the plane-wave background,
$\ket{0}$ is related to $|\rv \rangle$ as follows (for example, for
positive $p^+$):
\beq\label{0v}
|0\rangle = \theta^5_0\,\theta^6_0\,\theta^7_0\,\theta^8_0
|\rv \rangle~ .
\eeq
This, together with~(\ref{theta-flip}), implies that $\ket{0}$ and
$\ket{\rv}$ have opposite $Z_2$ parity. When $\mu\not=0$, $\ket{\rv}$
is the real vacuum state and thus it should be taken as $Z_2$
invariant. With this definition the world--sheet
action~\cite{Metsaev:2001bj} and the free string spectrum are also $Z_2$
invariant. At the interacting level, the $3$--string vertex should
also respect this $Z_2$ symmetry. By this we mean that two physical
amplitudes related by a $Z_2$ transformation should be exactly equal,
as it is for amplitudes 
that are connected by $SO(4) \times
SO(4)$ rotations. However, the simplest generalization of the
flat--space construction~\cite{Green:1982tc,Green:hw} to the plane--wave
background~\eq{pwave-back} considered in~\cite{Spradlin:2002ar} does
not satisfy this property: the physical amplitudes derived from the
vertices~\cite{Spradlin:2002ar,Pankiewicz:2002gs}
are $SO(4) \times SO(4)$ invariant, but transform non--trivially under
the $Z_2$ map~(\ref{Z2}). In fact, consider the vertices constructed
in~\cite{Spradlin:2002ar,Pankiewicz:2002gs}. They
all have the same structure
\be \label{H3-flat} 
\ket{H_3} = [\cdots]\; \ket{0}_{123}, \ee
where $[\cdots ]$ is an $SO(4)\times SO(4)\times Z_2$ invariant
operator, and we have denoted $\ket{0}_{123}:= |0\rangle_1 \otimes
|0\rangle_2 \otimes |0\rangle_3$ for convenience.  
For instance, one can compute the on-shell amplitudes
\be \label{Aij}
A^{IJ}  :=  \left({}_3\langle \rv| \a_{n (3)}^{I}\a_{-n (3)}^{J} \otimes
{}_2\langle \rv| \otimes {}_1\langle \rv| \a_{m (1)}^{I}\a_{-m
  (1)}^{J} \right)\; |H_3 \rangle.
\ee
One can then show that the amplitudes $A^{IJ}$ derived from the
vertices of the form~\eq{H3-flat} are not invariant under $Z_2$. This
can be done either by explicit computation of some
examples~\cite{Kiem:2002xn} or by using the $Z_2$ action directly
in~(\ref{Aij})~\cite{Chu:2002eu}. In particular, one finds that
$A^{ij} = -A^{i'j'}$. The basic reason for this asymmetry is that
either the $3$--string vertex~(\ref{H3-flat}) or the string
vacuum~(\ref{def-v}) are odd under $Z_2$ and thus the interacting
theory derived from~(\ref{H3-flat}) does not realize the $Z_2$
symmetry explicitly.

\subsection{Constructing the 3-string vertex}
 
In the covariant quantization the $3$--string vertex is basically
determined by its transformation properties under the BRST charge. In
the light--cone gauge all the world--sheet symmetries are fixed and
one has to follow a different method.
Following~\cite{Cremmer:1974jq,Green:1982tc,Green:hw} one can use the
space--time symmetries of the theory to fix the light--cone string
interaction. The construction consists of two steps.  First, one looks
for a string vertex $\ket{V}$ realizing locally on the world--sheet
all the kinematical symmetries of the light--cone algebra. Then one
has to add a particular polynomial prefactor term~\cite{Green:hw} in
order to respect also the dynamical part of the supersymmetry
algebra. In this note, we will concentrate on the kinematical
constraints, which can be satisfied with an ansatz where the bosonic
and the fermionic sector are factorized:
\be \label{V-pp}
\ket{V}= \delta\left(\sum_{\rr=1}^3 \a_\rr\right) 
\ket{E_a}\, \ket{E_b}\, ~.
\ee 
Here $\a_\rr$ is related to the $+$ component of the string momentum
($\a_\rr =2 p^+_\rr$) and $\ket{E_a}$ (resp. $\ket{E_b}$) is the
contribution from the bosonic (resp. fermionic) oscillators. Moreover, $\ket{E}$
and $\ket{V}$ are kets in the tensor product of the three independent
Hilbert spaces describing the external strings. They have the same
structure containing a bilinear exponential part acting on the
vacuum\footnote{
In the flat space case the vertex $\ket{V}$ has a slightly different
structure, since the zero--mode bosonic momenta 
$\hat{p}_{0(\rr)}$ and
fermionic momenta $\hat{\l}_{0(\rr)}$ have continuous spectra. In the
pp--wave case, both $\ph_{0(\rr)}$ and $\lh_{0(\rr)}$ are rewritten in
terms of oscillators and are not different from the 
nonzero modes.}. For
instance $\ket{E_a} = \exp{(\sum \ah_{n (\rr)}^{\dagger} N^{\rr
\rs}_{nm} \ah_{m (\rs)}^{\dagger} ) } \ket{0}_{123}$. In this formalism the
kinematical constraints become
\bea
&&\sum_{\rr=1}^3 \sum_{n \in {\rm\bf Z} } \a_{\rr} X^{(\rr)}_{mn}
\hat{x}_{n(\rr)}  \ket{E_a} =0, 
\quad 
 \sum_{\rr=1}^3 \sum_{n \in {\rm\bf Z} } X^{(\rr)}_{mn}
\hat{p}_{n(\rr)} \ket{E_a} =0, 
\label{c1} \\ 
&& \sum_{\rr=1}^3 \sum_{n \in {\rm\bf Z} } \a_\rr X^{(\rr)}_{mn}
\hat{\th}_{n(\rr)}  \ket{E_b} =0, \quad 
\sum_{\rr=1}^3 \sum_{n \in {\rm\bf Z} } X^{(\rr)}_{mn}
\hat{\l}_{n(\rr)} \ket{E_b} =0~, \label{c2} 
\eea
where we added the hats on the various modes to stress that they are
operators acting on the string Hilbert spaces. Notice that all these
constraints (anti)--commute among themselves, thanks to the identity
$\sum_{r=1}^3\a_r (X^{(r)}X^{(r)T})_{mn}=0$. Thus there is hope to
find a state $\ket{V}$ satisfying all the
eqs.~\eq{c1}--\eq{c2}. However, it is a challenging task to find
the vertex by direct solution of the above constraints. The idea
of~\cite{Cremmer:1974jq} is to write an ansatz for $\ket{V}$ in an
integrated form where one can show that the constraints hold, and then
derive $\ket{V}$ in the oscillator space by performing
explicitly the integrals. In the bosonic sector this procedure
completely fixes the form of the matrix $N^{\rr \rs}_{nm}$ appearing in the
exponential. In this case the integrated ansatz is
\beq
\ket{E_a} =  
\prod_m \delta\left( \sum_{\rr=1}^3 \sum_{n \in {\rm\bf Z} }
X^{(\rr)}_{mn} \hat{p}_{n(\rr)}\right)
\int [dp] \;\prod_{\rr=1}^3 {\prod_{{k= -\infty}}^\infty}
\psi(\ah^\dag_{k (\rr)}, p_{k (\rr)}) \ket{\rv}_{123}.
\label{Ea}
\eeq 
Here the $\hat{p}$'s are the operators in the string mode expansion,
while the $p$'s are just $c$--numbers and are integrated with the
measure $[dp] = \prod_{\rr=1}^3 \prod_{k= -\infty }^\infty dp_{k
(\rr)}$. The operators $\psi(\ah^\dag_{k},p_{k})$ are related to the
harmonic oscillator wave-function of a state with occupation number
$k$ (see, for instance, eq.~(3.3) of~\cite{Pankiewicz:2002gs}).  It is
easy to show that~\eq{Ea} satisfies both conditions~\eq{c1}.  In order
to prove the validity of the second constraint~\eq{c1}, one exploits
the fact that the ket $\psi(\ah^\dag_{k},p_{k}) \ket{\rv}$ is an
eigenvector of the momentum operator $\hat{p}_{k}$. Thus one can
insert the $\delta$--function in the integral and eliminate all the
hats both in the integrand~\eq{Ea} and in the constraint. The integral
becomes then of the form $\int dx\;x\, \delta(x)$ and is clearly
vanishing. On the contrary, for the first equation in~(\ref{c1}), it
is easier to keep the $\delta$--function out of the integral and
commute the constraint inside the integration. Then, one can realize
the operators $\hat{x}$ as $\partial/\partial {p}$ and obtain an
integrand that is a total derivative. The boundary terms do not
contribute because of the Gaussian factor $\exp{(-p_{k (\rr)}^2)}$ in
the $\psi(\ah^\dag_{k},p_{k})$.

The generalization of this procedure to the fermionic sector is
straightforward, except for a subtle point in the treatment of the
zero modes. The fermionic analogues, $\chi(\bh^\dag_{\pm k},\l_{\pm k})$,  
of the bosonic operators $\psi(\ah^\dag_{k},p_{k})$  can
only be defined if the fermionic oscillators are paired.  The
non--zero mode creation operators can be naturally paired
($\bh^\dag_{k},\bh^\dag_{-k}$), but the oscillators $\bh^\dag_{0}$ have to
be treated separately. So the form of $\ket{E_b}$ satisfying~\eq{c2}
for $m=0$ has to be supplied by hand to the integrated ansatz. Thus we
first look for a state satisfying simultaneously
\beq\label{fcon}
\sum_{\rr=1}^3 \lh_{0 (\rr)}^a |\delta\rangle = 0 ~~,\qquad
\sum_{\rr=1}^3 \a_{\rr} \thh_{0 (\rr)}^a |\delta\rangle = 0 ~,
\eeq
In flat space, these conditions are usually solved by the
following state 
\be\label{fszms}
\ket{E_b^0} = \prod_{a=1}^8
\left(\sum_{\rr=1}^3 \lh_{0(\rr)}^a\right)  \ket{0}_{123}.
\ee
In~\cite{Spradlin:2002ar} and in many subsequent papers this same zero
mode structure has been adopted also in the construction of the string
interaction vertex in the background~\eq{pwave-back}. However from the
discussion of the previous section, it is clear that eq.~\eq{fszms}
has a quite different behaviour in flat space and in the
plane--wave background: in the first case one can define both the vacuum of
the theory $\ket{0}$ and eq.~\eq{fszms} to be $SO(8)$ invariant, while
in the second case either the true vacuum $\ket{\rv}$ or
eq.~\eq{fszms} are $Z_2$--odd. However as explained in
\cite{Chu:2002eu}, there is a different solution of the
constraints~\eq{fcon} which is $Z_2$ symmetric together with the definition
$Z_2 \ket{\rv} = \ket{\rv}$
\beq\label{newdelta}
|\delta\rangle = \prod_{a=1}^8
\left(\sum_{\rr=1}^3 \lh_{0(\rr)}^a\right)
\prod_{a=1}^8 \left(\sum_{\rr=1}^3 \a_{\rr} \thh_{0 (\rr)}^a\right)
\ket{\rv}_{123}~. 
\eeq

In the appendix C of \cite{Pankiewicz:2002gs}, it is claimed that the
zero mode delta-function $\ket{\d}$ cannot be extended to include
non--zero modes such that \eq{c2} are satisfied. Of course, if the
zero mode structure $\ket{\d}$ is inserted in the same integral ansatz
used in flat space, the first condition in~\eq{c2} is not
satisfied. However, we want to stress here that the form of the
integral ansatz usually employed in flat space does not have a
fundamental meaning. Its main virtue is to exploit a physical
requirement like momentum conservation to solve the kinematical
constraints, but this ansatz does not have to be valid in all
backgrounds and can be modified to satisfy possible additional
symmetry requirements. In the plane--wave case, it is not difficult to
deform the usual answer for flat space and to write an exponential
vertex that satisfies all the constraints~\eq{c1} and~\eq{c2}. For
example,
\bea\label{E-la}
\ket{E_b} & = & {\rm exp}\Big[\sum_{\rr,\rs=1}^3\sum_{m,n=1}^{\infty} 
\bh^{\dagger}_{-m(\rr)} \bar{Q}_{mn}^{\rr\rs} \bh^{\dagger}_{n(\rs)}
- \sqrt{2} \Lambda \sum_{\rr=1}^3\sum_{m=1}^{\infty} 
\bar{Q}^\rr_m \bh^{\dagger}_{-m(\rr)} \Big]' \\ \nonumber
& \times & {\rm exp}\Big[\sum_{\rr,\rs=1}^3\sum_{m,n=1}^{\infty} 
\bh^{\dagger}_{m(\rr)} \bar{Q}_{mn}^{\rr\rs} \bh^{\dagger}_{-n(\rs)}
+ \frac{\alpha}{\sqrt{2}} \Theta \sum_{\rr=1}^3\sum_{m=1}^{\infty} 
\bar{Q}^\rr_m \bh^{\dagger}_{m(\rr)} \Big]''~,
\eea
where 
\be
\L : = \a_1 \lh_{0(2)} - \a_2 \lh_{0(1)}, \quad 
\Theta:=  \frac{1}{\a_3} (\thh_{0(1)} - \thh_{0(2)})
, \quad \a = \a_1\a_2\a_3~.
\ee
$[\cdots ]'$ denote a summation over the positive
$\Pi$--chirality components ($a=1,\cdots, 4$) of the spinors
$b^{\dagger a}$, while $[\cdots ]''$ denote a summation over the
negative $\Pi$--chirality components ($a=5,\cdots, 8$). 
The matrices $Q$ are diagonal in the spinor space and we have suppressed
the spinor indices. In the notation of~\cite{Pankiewicz:2002gs} they read
\bea
&&\bar{Q}^{\rr\rs}_{mn}:= e(\a_\rr) \sqrt{\frac{|\a_\rs|}{|\a_\rr|}}\;
[U_{(\rr)}^{1/2} C^{1/2} N^{\rr\rs} C^{-1/2}
U_{(\rs)}^{1/2}]_{mn},\label{defQ1} \\ 
&&\bar{Q}^{\rr}_m := \frac{e(\a_\rr)}{\sqrt{|\a_\rr|}}
[U_{(\rr)}^{1/2} C_{(\rr)}^{1/2} C^{1/2}
N^\rr]_m~\label{defQ2}.
\eea
In the Appendix we explicitly show that the exponential~\eq{E-la}
satisfies the constraints~\eq{c1} and \eq{c2}. In this respect the
vertex presented here is on the same footing as the vertex proposed
in~\cite{Pankiewicz:2002gs}. However for the vertex~\eq{E-la}, the
symmetry under the full $SO(4) \times SO(4) \times Z_2$ is compatible
with the invariance of the vacuum $\ket{\rv}$, while this is not true
for the vertex in~\cite{Pankiewicz:2002gs}, which is 
built on the zero--mode structure~\eq{fszms}. Thus,
just like the other generators of the spacetime symmetry algebra, the
discrete $Z_2$ symmetry plays an important role in fixing the form of
the string interaction. In fact, it can be used to distinguish between
different forms for the kinematical part of the interaction (the one
of~\cite{Pankiewicz:2002gs} and the one in eq.~\eq{E-la}) which
otherwise have the same properties.
In our opinion, only after having correctly implemented this discrete
symmetry at the level of the kinematical vertex, it is possible to
consider the dynamical symmetries and to look for the supersymmetric
completion of the vertex by determining the prefactor.

\section{Discussion}

The main result of this note is to adapt the usual construction of the
light--cone string interaction to the plane--wave case. In particular, we
showed that it is actually possible to construct a $3$--string vertex
that satisfies all the kinematical constraints and gives, at the same
time, $Z_2$--invariant on--shell amplitudes. Since we are considering
a {\em discrete} symmetry, the transformation properties of the
various physical quantities can not depend on any continuous
parameter. Thus the $Z_2$--symmetry should appear also in the
perturbative Yang--Mills computations, which are valid in the 
large $\mu$ limit. However, while this symmetry is manifest in the string
setup, its realization on the  Yang--Mills side is 
much less understood. At the present the only explicit computation with
operators mixing the two $SO(4)$'s is the one
of~\cite{Gursoy:2002yy}. In that analysis, it turns out that the
results involving some two impurity operators are surprisingly (from
the field theory point of view) symmetric. In~\cite{Gursoy:2002yy}
this behaviour was explained by arguing that the two impurity operators
considered were connected by a supersymmetry transformation. This idea
has been expanded in various places in the recent literature and it
has been proposed that all the 256 two impurity operators are
connected by the supersymmetry transformations generated by the 16
supercharges commuting with the light--cone Hamiltonian  
\footnote{See, for example, the version 2 of the second paper 
in~\cite{Beisert:2002bb}.}. Here we
just notice that this observation is nicely consistent with the
$3$--string vertex presented in this note, since it yields more
symmetric result among some two impurity operators. On the contrary,
all the vertices of the form~\eq{H3-flat} imply $A^{ij} = -A^{i'j'}$
and $A^{ij'} = 0$, which is a puzzling result if all the two impurity
states are in the same long supermultiplet.

Of course, the $Z_2$ invariance of the Yang--Mills results needs to be
more thoroughly tested. At the moment a systematic approach is quite
difficult since we do not yet have a clear and general recipe to
compare string and field theory computations. Moreover, recently there
has been a radical change of perspective in the plane-wave/CFT
correspondence. 
In the early dynamical comparisons between string and
field theory, the idea was to keep valid, also at the interacting
level, the dictionary between string states and YM operators proposed
in~\cite{Berenstein:2002jq}. In this framework each computation
represents an independent test of the duality. 
However more recently it has
been proposed that the dictionary has to be adjusted in order to have
on the Yang--Mills side orthogonal operators~\cite{Beisert:2002bb}.
Of course there are many different
field theory basis satisfying  this requirement.
A particular ``string theory'' basis on the YM side has been singled out 
using results of the $3$--string vertex.
Clearly, if this approach is correct, the results of $3$--string
amplitudes do not always represent an independent check of the duality.

Another dual description of string theory in the plane--wave background
is the string bit model proposed in~\cite{Verlinde:2002ig}. In this
framework, already now it is possible to deal with a larger class of
two impurity operators. 
In fact the string bit results
of~\cite{Vaman:2002ka} are valid for all the states with two
bosonic oscillators. Of course, also in this context one has 
the problem of fixing the dictionary between the spectra of
string theory and of string bits. However, the symmetry properties of
the $3$--point interaction do not seem to be 
sensitive to this problem;
and quite interestingly it seems that the string bit 
interactions are
$Z_2$ symmetric. If this is correct, the agreement between
the interaction vertex of~\cite{Spradlin:2002ar} and the string bit
computations (see the second paper in~\cite{Vaman:2002ka}) 
cannot be extended to the bosonic operators with both vector indices in
the second $SO(4)$. This suggests that the $Z_2$ transformation
properties of the amplitudes can represent a check of the relation
among the various dual descriptions that can be done without the need
of fixing the precise dictionary between operators and states.
Of course
it would be very interesting to see whether this pattern also appears 
in the computation on Yang--Mills side of the correspondence.

\section*{Acknowledgments}

We would like to thank the organizers and participants of the 35th
Symposium Ahrenshoop ``Recent Developments in String/M-Theory and Field
Theory Symposium Ahrenshoop", Berlin, Aug 26-30, 2002, and of the workshop 
``The quantum structure of space-time and the geometrical nature of
fundamental interactions'', Leuven, September 13-19 2002. We would
also like to thank Paolo Di Vecchia, Valentin V. Khoze, Feng-Li Lin
and Jens Lyng Petersen for interesting discussions. This work is
supported in part by EU RTN contracts HPRN-CT-2000-00122 and
HPRN-CT-2000-00131, Nuffield foundation of UK and the National Science
Council of Taiwan. MP, RR and AT are supported by European Commission
Marie Curie Postdoctoral Fellowships.

\section*{Appendix A: The fermionic constraints}

In this appendix we will show that the matrices  
$\bar{Q}^{\rr\rs}_{mn}$ and $\bar{Q}^{\rr}_{m}$ do indeed
satisfy the fermionic kinematical constraints of eq.~(\ref{c2}).
For $m=0$ the constraints reduce to eq.~\eq{fcon}  
and are automatically solved by
the way the zero mode part of the vertex is defined.
Notice that for $m\not= 0$ the zero modes appear in \eq{c2} only
through the 
combinations $\Theta$ and $\Lambda$, which anticommute with the zero
mode structure \eq{newdelta}. Thus the way 
to proceed is to get rid of the annihilation operators in $\lh_{n}$ and
$\thh_{n}$, by commuting them through the exponential part of the vertex and 
reducing to a set of equations involving only the operators 
$\bh^{\dagger}_{n (\rr)}$, $\Theta$ and $\Lambda$. The
coefficients of such operators then provide the final constraint equations.
Since the matrices $\bar{Q}$'s appearing in the vertex are
diagonal in the spinor space, we can write a set of equations holding
for both positive and negative $\Pi$--chirality spinor (namely for each
fixed index $a=1,\ldots,8$).
In the notation of \cite{Pankiewicz:2002gs}, we have: 

\underline{$m>0$} 
\begin{eqnarray}
&& B+\sum_{\rr=1}^3e(\a_{\rr})\sqrt{|\a_{\rr}|}A^{(\rr)}
C_{(\rr)}^{-1/2}U_{(\rr)}^{1/2}\bar{Q}^{\rr} =  0\,,\label{1p}\\
&& \sqrt{|\a_{\rs}|}A^{(\rs)}C_{(\rs)}^{-1/2}U_{(\rs)}^{-1/2}+
\sum_{\rr=1}^3e(\a_{\rr})\sqrt{|\a_{\rr}|}A^{(\rr)}
C_{(\rr)}^{-1/2}U_{(\rr)}^{1/2}\bar{Q}^{\rr \rs} =  0\,,\label{2p}\\
&&\sqrt{|\a_{\rs}|}A^{(\rs)}C_{(\rs)}^{-1/2}U_{(\rs)}^{1/2}-
\sum_{\rr=1}^3e(\a_{\rr})\sqrt{|\a_{\rr}|}A^{(\rr)}
C_{(\rr)}^{-1/2}U_{(\rr)}^{-1/2} \bar{Q}^{\rs \rr}\,{}^T + 
\a B\bar{Q}^{\rs\,T}=0.
\label{3p}
\end{eqnarray}
For the positive (resp. negative) $\Pi$--chirality spinors  
the first two equations come from the constraint on $\lh_{n}$ (resp. $\thh_{n}$) 
in (\ref{c2}), while
the third one comes from the equation involving $\thh_{n}$ (resp. $\lh_{n}$).

\underline{$m<0$} 
\begin{eqnarray}
&&\sum_{\rr=1}^3\frac{1}{\sqrt{|\a_{\rr}|}}A^{(\rr)}CC_{(\rr)}^{-1/2}
U_{(\rr)}^{-1/2}\bar{Q}^{\rr} =  0\,,\label{1n}\\
&& A^{(\rs)}CC_{(\rs)}^{-1/2}-e(\a_{\rs})\sqrt{|\a_{\rs}|}
\sum_{\rr=1}^3\frac{1}{\sqrt{|\a_{\rr}|}}
A^{(\rr)}CC_{(\rr)}^{-1/2}U_{(\rr)}^{-1/2}\bar{Q}^{\rr \rs}U_{(\rs)}^{-1/2} = 0\,,
\label{2n}\\
&& A^{(\rs)}CC_{(\rs)}^{-1/2}+e(\a_{\rs})\sqrt{|\a_{\rs}|}
\sum_{\rr=1}^3\frac{1}{\sqrt{|\a_{\rr}|}}
A^{(\rr)}CC_{(\rr)}^{-1/2}U_{(\rr)}^{1/2}\bar{Q}^{\rs \rr}\,{}^T 
U_{(\rs)}^{1/2}
= 0\,.\label{3n}
\end{eqnarray}
In this case the situation is reversed. Eqs.~(\ref{1n}) and (\ref{2n})  
descend from the $\thh_{n}$ 
constraint for the positive $\Pi$--chirality modes and 
from the $\lh_{n}$ one for the negative modes. 
Vice--versa, the third equation comes from $\lh_{n}$ 
for the positive $\Pi$--chirality modes and from $\thh_{n}$
for the negative ones.

The proof for the non-zero mode constraints is rather tedious 
but straightforward. To give
an idea of how it works we will explicitly solve the constraint in
eq.(\ref{3p}). Inserting in the second term of eq.(\ref{3p}) the 
expression for $\bar{Q}^{\rr \rs}$, we obtain

\begin{equation}
\sum_{\rr=1}^3 e(\a_\rs) \frac{\a_{\rr}}{\sqrt{|\a_{\rs}|}}A^{(\rr)}
C_{(\rr)}^{-1/2}C^{-1/2} N^{\rr \rs} C^{1/2} U_{(\rs)}^{1/2}.
\label{proof1}
\end{equation}
Then using the identity (this and similar identities follow from 
imposing the bosonic constraints, see for instance
\cite{Pankiewicz:2002gs})
\begin{equation}
\sum_{\rr=1}^3\a_\rr A^{(\rr)}C_{(\rr)}^{-1/2}
C^{-1/2} N^{\rr \rs}C^{1/2}
=  \a B\left[C_{(\rs)}^{1/2}C^{1/2} N^\rs\right]^T 
+ \a_\rs A^{(\rs)}C_{(\rs)}^{-1/2},
\end{equation}
and the definition of $\bar{Q}^\rr_m$, eq.(\ref{defQ2}), we can
rewrite eq.(\ref{proof1}) as
\begin{equation}
\sqrt{|\a_{\rs}|}A^{(\rs)}C_{(\rs)}^{-1/2}U_{(\rs)}^{1/2}
+\a B\bar{Q}^{\rs\,T},
\end{equation}
which cancel the other two terms in eq.(\ref{3p}).

\end{document}